\documentclass[prb,superscriptaddress,showpacs,papersize=a4paper,notitlepage]{revtex4-1}%
\usepackage[pdftex]{graphicx}
\usepackage{bm}
\usepackage{etoolbox}
\usepackage[usenames,dvipsnames]{color}
\usepackage{mathtools}
\usepackage{amsmath}
\usepackage{amsfonts}
\usepackage{amssymb}

\begin{document}

\title{Perturbed Kitaev model: excitation spectrum and long-ranged spin correlations.}
\author{A. V. Lunkin}
\affiliation{Skolkovo Institute of Science and Technology, 143026 Skolkovo, Russia}
\affiliation{ L. D. Landau Institute for Theoretical Physics, Kosygin str.2, Moscow
119334, Russia}
\affiliation{ Moscow Institute of Physics and Technology, Moscow 141700, Russia}
\author{K. S. Tikhonov}
\affiliation{ L. D. Landau Institute for Theoretical Physics, Kosygin str.2, Moscow
119334, Russia}
\affiliation{National University of Science and Technology "MISiS", 119049 Moscow, Russia}
\author{M. V. Feigel'man}
\affiliation{ L. D. Landau Institute for Theoretical Physics, Kosygin str.2, Moscow
119334, Russia}
\affiliation{ Moscow Institute of Physics and Technology, Moscow 141700, Russia}

\begin{abstract}
We developed general approach to the calculation of power-law infrared asymptotics of spin-spin correlation functions
in the Kitaev honeycomb model with different types of perturbations.  We have shown that in order to find these
correlation functions, one can perform averaging of some bilinear forms composed out of free Majorana fermions,
and we presented the method for explicit calculation of these fermionic densities.
We demonstrated how  to derive an effective Hamiltonian for the Majorana fermions,
including the effects of perturbations.  For  specific application of the general theory, we have studied the effect 
of the  Dzyaloshinskii-Moriya anisotropic spin-spin interaction; we demonstrated that it leads, already 
in the second order  over its relative magnitude $D/K$, to a power-law spin correlation functions, 
and calculated dynamical spin structure factor of the system.
We have shown that an external magnetic field $h$ in presence of the DM interaction,
 opens a gap in the excitation spectrum of magnitude $\Delta \propto D h$.
\end{abstract}

\maketitle


\section{Introduction}
Quantum spin liquids (QSL) (see e.g. Refs.~\onlinecite{PWA1,PWA2,Wen,rev1,rev2,balents2010spin,savary2016quantum}) present examples of strongly correlated quantum phases not developing any kind of local order in spite of vanishing specific entropy at zero temperature.  \emph{Critical}, or algebraic QSL's are characterized by spin correlation functions decaying with distance (time) as certain powers. An exactly solvable case of the critical QSL is provided by the celebrated Kitaev honeycomb spin model~\cite{Kitaev06} (see also Ref. \onlinecite{nussinov2013}). This model was originally invented as a simplest solvable spin model possessing nontrivial topological phases, relevant in the context of topological quantum computing; 
later it has been  found that similar spin interactions can be realized in the honeycomb-lattice oxides\cite{jackeli09} Na$_2$IrO$_3$ and Li$_2$IrO$_3$. Kitaev honeycomb model Hamiltonian reads
\begin{equation}
\mathcal{H_K}=K\sum_{\left\langle ij\right\rangle \in \alpha\beta(\gamma) } \sigma_i^\alpha \sigma_j^\alpha
\label{Kitaev_Hamiltonian}
\end{equation}
where each of $\sigma_i^\alpha$ represents a Pauli matrices corresponding to  $i$-th site of the  honeycomb lattice. There are 3 types of bonds (see Fig. \ref{flux}) which are denoted below as $xy(z)$,$yz(x)$ and $zx(y)$. Although long-range spin correlations vanish exactly in the model Eq (\ref{Kitaev_Hamiltonian}), its spectrum contains gapless fermions and hence it presents a convenient starting point for the construction of controllable theories possessing long-range spin correlations. In a realistic situation, low-energy effective description of these materials is given by a mixture of the Kitaev and Heisenberg (and other) interactions with weights depending on the microscopic parameters\cite{PhysRevLett.110.097204,winter2016challenges} (for recent reviews on "Kitaev materials" see Refs.~\onlinecite{trebst2017kitaev,winter2017models}.). Another interesting appearance of Heisenberg-Kitaev (HK) model is in a low-energy theory of a Hubbard model on a honeycomb lattice with spin-dependent hopping\cite{hassan13}. An important question is to which extent the properties of the ground state (and excitations) of the perturbed Kitaev model are proximate to that of the unperturbed one? Exact diagonalization and a complementary spin-wave analysis\cite{chaloupka10} show that spin-liquid phase near the Kitaev limit is stable with respect to small admixture of Heisenberg interactions\cite{PhysRevB.84.100406,shaffer12}. From experimental side, although there was no report of detection of a spin-liquid state realized by this scenario in the absence
of external magnetic field, the importance of the Kitaev interaction has been clearly demonstrated in $\alpha$-RuCl$_3$\cite{banerjee2016proximate,banerjee2017neutron,PhysRevB.90.041112,PhysRevLett.114.147201,PhysRevB.91.241110,janvsa2017observation} and Na$_2$IrO$_3$ \cite{chun2015direct,choi2012spin,PhysRevLett.114.017203}. Moreover, in two recent Refs. \onlinecite{PhysRevLett.119.037201} and \onlinecite{hentrich2017large} the data are present in favor of a spin liquid state 
in $\alpha$-RuCl$_3$ realized upon application of external magnetic field.

As we have already mentioned, Kitaev model in its original form does not possess long-range spin correlations; moreover, its spin correlators 
are strictly local\cite{Baskaran07}. A perturbative addition of the  Heisenberg interaction  does not change this fact\cite{baskaran11}. In order to produce a spin-liquid phase  with long-range correlations, some other terms should be added to the effective Hamiltonian.  The simplest perturbation which does not destroy spin-liquid phase but renders correlations non-local is magnetic field\cite{TFK2011} (the HK model  with magnetic field  was also studied in Refs \onlinecite{horsch11, jiang2011possible}). A more general picture of emergence of spin correlations in the Kitaev model due to various perturbation was discussed recently in Ref. \onlinecite{PhysRevLett.117.037209}. In the present paper, we develop a general scheme in which the long-distance (time) spin-spin correlation functions in Kitaev honeycomb model with various types of local perturbations can be analysed. In quite a general form, we have reduced the calculation of spin-spin correlation functions to evaluation of specific correlation functions bilinear forms of free Majorana fermion. We have also studied effects of higher-order terms and found the effective Hamiltonian for Majorana fermions modified by weak perturbations of the spin Hamiltonian. Besides explaining qualitative differences between results previously obtained for various perturbations\cite{TFK2011,PhysRevLett.117.037209}, we considered the effect of Dzyaloshinksi-Moria (DM) interaction and demonstrated that it leads to the power-law correlations already in the lowest possible (second)
order over its strength $D/K$. We have also found that application of magnetic field $\mathbf{h}$ to a model, perturbed by DM interaction, produces the gap $\Delta \sim D h/K$ in the Majorana fermion spectrum, rendering decay of spin correlations in space and time exponential.  Note in this respect that linear growth of the spin-liquid excitation gap has been recently observed\cite{PhysRevLett.119.037201,hentrich2017large} in $\alpha$-RuCl$_3$.


\section{Kitaev honeycomb model with perturbations: a brief review}

The model (\ref{Kitaev_Hamiltonian}) was solved \cite{Kitaev06} via an exact transformation to  Majorana fermion representation (similar procedure using the language of Jordan-Wigner transformation was later developed in Ref. \onlinecite{chen2008exact}), which is constructed as follows.  For each lattice site $i$ one defines  4 Majorana fermions 
$b^x_i$, $b^y_i$, $b^z_i$ and $c_i$.  Our goal is to reproduce the algebra of Pauli matrices using these fermions, which
is achieved by the substitution $\sigma_i^\alpha=ib^\alpha_i c_i$.  However, the number of degrees of freedom
is larger in the Majorana representation (4 states per site) than in the spin-$\frac12$ representation: each pair of  Majorana fermions is equivalent
 to one usual complex fermion, so to each site two fermions are attached, which is equivalent to  2 spin-$\frac{1}{2}$ variables. 
This problem is fixed by the constraint imposed on the Majorana operators while acting in the "physical subspace"\,
of the full  Hilbert space of the Majorana operators: $b^x_i b^y_i b^z_i c_i=1$. This condition is due to
the identity $\sigma^x \sigma^y \sigma^z=i$ valid for Pauli matrices. 

\begin{figure}
\includegraphics[scale=0.5]{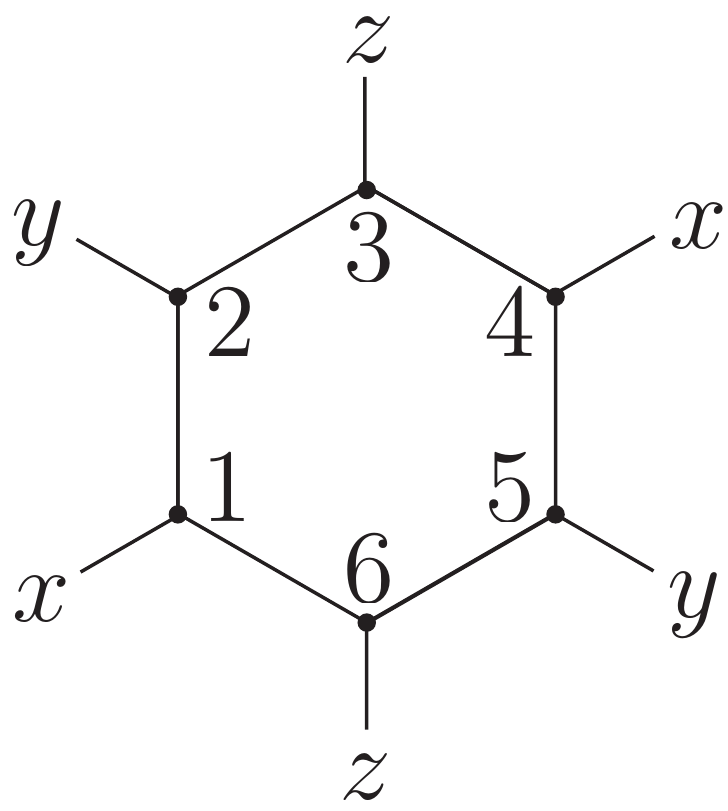}
\caption{Definition of gauge-invariant integrals of motion $W_p=u_{21}u_{61}u_{65}u_{45}u_{43}u_{23}=\sigma^x_1\sigma^y_2\sigma^z_3\sigma^x_4\sigma^y_5\sigma^z_6$.}
\label{flux}
\end{figure}

 After the substitution of the spin operators in terms of Majorana variables, the Hamiltonian becomes
\begin{eqnarray}
H=-iK\sum_{\langle i,j\rangle \in \alpha\beta(\gamma) } b_i^\gamma b_j^\gamma c_i c_j
\end{eqnarray}
The honeycomb lattice contains two sublattices and each edge connects vertices from different sublattices.  Let us take
as convention that for each bond of the lattice  in the above sum  $i\in$ the first sublattice and $j\in$  the second one. Then for each of the bonds   $\langle i,j \rangle \in \alpha\beta(\gamma)$ there exists a conserved quantity 
$u_{ij}=i b^\gamma_i b^\gamma_j$:  such an operator commutes with the Hamiltonian.
The operator $u_{ij}$ also  anti-commutes with $\sigma^\gamma_{i}$ or $\sigma^\gamma_j$, and  also $u_{ij}^2=1$,
thus  $u_{ij}=\pm1$ for all eigenstates. As we have already mentioned above,  the Majorana representation has extra
 degrees of freedom w.r.t.  the original  Pauli matrix representation. These extra degrees of freedom are partially accounted for  by the  constraints $b^x_i b^y_i b^z_i c_i=1$.
Another extra degree of freedom is related to the gauge transformation: for any  particular site  $i$  we  can replace $b^\alpha_i\rightarrow -b^\alpha_i$ and $c_i\rightarrow -c_i$;
in result, all $u_{ij}$, where $j$ is a neighbour of $i$, change its sign.  Gauge independent integrals of motion, called "fluxes", are defined for all plaquettes of  the lattice, see Fig. \ref{flux}.  For each plaquette $p$  a flux operator is defined as
$w_p=\prod_{\langle i,j \rangle \in p } u_{ij}$ (\ref{flux}).  Evidently, $w_p^2=1$, thus  $w_p=\pm1$ for all eigenstates.
 If some $w_p=-1$,  we say that there is a flux associated with  the $p$-th  plaquette.  It can be shown\cite{lieb2004flux} that the ground state contains  no fluxes.  Thus, up to possible  gauge transformations,  in the ground state all 
$u_{ij}=1$. As a result, the Majorana  Hamiltonian can be written in the following form:
\begin{eqnarray}
H=-iK\sum_{\langle ij\rangle} c_i c_j
\label{free_hamiltonian}
\end{eqnarray}

Fourier transformation of the Hamiltonian (\ref{free_hamiltonian}) leads to the free-particle  spectrum $\varepsilon(\mathbf{p})=|f(\mathbf{p})|$ where $$f(\mathbf{p})=2K(e^{i(\mathbf{n}_1,\mathbf{p})}+e^{i(\mathbf{n}_2,\mathbf{p})}+1)$$
 and $ \mathbf{n}_1=\left(\begin{smallmatrix}\frac{1}{2}\\ \frac{\sqrt{3}}{2}\end{smallmatrix}\right)$ and $\mathbf{n}_2=\left(\begin{smallmatrix}-\frac{1}{2}\\ \frac{\sqrt{3}}{2}\end{smallmatrix}\right)$ are the translation vectors of the lattice
(lattice constant is set to unity). Here  $x$ axis is perpendicular the $z$-bonds, while  $y$ axis is parallel the $z$-bonds (see Fig. \ref{flux}). This  energy spectrum possesses
two conical points: $\mathbf{ Q}_{1,2}=\left(\begin{smallmatrix}
\pm \frac{2\pi}{3} \\
\frac{2\pi}{\sqrt{3}} 
\end{smallmatrix}\right )
$. 
Near these points $\varepsilon(\mathbf{p})$ vanishes linearly with $|\mathbf{p}-\mathbf{ Q}_{1,2}|$.

For  a system with a spectrum containing conical points, one would  expect a power-law decay of  correlation functions
at long times and large space separations.  However,  it is not true for the Kitaev model as was highlighted in 
Ref.~\onlinecite{Baskaran07} demonstrated that spin correlations vanish for any distance $i-j$ longer than just 
single bond length.  Such a strange behavior originates from exact integrability of the model.  
Indeed, any spin operator $\sigma^\alpha_i$ anticommutes with all bond  variables $u_{ij}$;  as a result, the action
of a spin operator on the flux-less ground state  is to create two fluxes located in the plaquettes $p_{1,2}^{i,j}$
adjacent to the bond $(ij)$.  In order for the correlation function 
$\langle T \sigma_i^\alpha(t) \sigma_j^\beta(0) \rangle$  to be nonzero,  fluxes created by the spin operator
$\sigma_i^\alpha $ must be annihilated by another operator $\sigma_j^\beta $, since overlap between quantum states
with different flux content is zero due to conservation of the fluxes (integrability). The above condition
for flux cancellation cannot be fulfilled for  $|i-j| > 1$, leading to vanishing of such spin correlations.
It is evident from the above discussion that small perturbations,
 which weakly destroy integrability of the Kitaev model, allow for the "revival" \, of spin correlations at large
distances.  The type of these correlations (exponential vs power-law decay~\cite{baskaran11,TFK2011})
depends upon the flux content of the perturbing operator, as we discuss below.

Let us first describe the relevant perturbations to the Kitaev Hamiltonian. As magnetic materials with heavy elements contain strong spin-orbit interactions responsible for establishing the interaction of the form of Eq. (\ref{Kitaev_Hamiltonian}),
it is necessary to classify the states in terms of total angular  momentum $K$, without separation into spin and orbital parts. In a number of cases low-energy excitations of such a system can be described in terms of \textit{effective} 
variables, formally equivalent to spin-$\frac{1}{2}$. In some magnetic  materials (for example, $Na_2IrO_3$, $\alpha$-RuCl$_3$, $\alpha$-Li$_2$IrO$_3$, see Ref. \onlinecite{winter2016challenges}) 
the anisotropic  Kitaev term is the largest in magnitude.  However, the full Hamiltonian contains also
other terms which can  possibly  be treated as perturbations to the Kitaev Hamiltonian.
Most typical perturbations  are known to be isotropic Heisenberg interaction and  pseudodipolar interaction:  
\begin{eqnarray}
V_H=\sum_{\langle i,j \rangle} J (\vec{\sigma_i},\vec{\sigma_j}) \quad \mbox{Heisenberg interaction } \label{heisenberg} \\
V_\Gamma=\sum_{\langle i,j \rangle \in \alpha\beta(\gamma)} \Gamma^\gamma_{ij}(\sigma^\alpha_i \sigma^\beta_j+\sigma^\alpha_j\sigma^\beta_i)\quad \mbox{ pseudodipolar interaction}
\end{eqnarray}
Many different ground-states can be realized~\cite{rau2014generic} depending upon specific relations between
energy parameters $K$, $J$ and $\Gamma$: ferromagnetic, anti-ferromagnetic, zigzag or stripy types of ordered
states, as well as spin-liquid state in some narrow range of the full parameter space. 

More detailed microscopic study of effective spin interactions, conducted in Ref. \onlinecite{winter2016challenges} demonstrated relevance of DM interactions.  In particular it is interesting to consider the effect of
the DM anisotropic interaction. It was found that DM interaction between nearest neighbours is absent in this system due to inversion symmetry of the
 microscopic Hamiltonian. Interaction between  \textit{next-nearest-neighbours}
\begin{eqnarray}
V_{DM}=\sum_{\langle \langle i,j \rangle\rangle} (\vec{D}_{ij},[\vec{\sigma}_i,\vec{\sigma}_j])
 \label{DM}.
\end{eqnarray}
is compatible with inversion symmetry. The honeycomb lattice possesses three different types of next-nearest-neighbours. 
Each of them can be put  into one-to-one correspondence with nearest-neighbours directions $\alpha$:  if the pair of next-nearest-neighbours
is connected by a segment which is perpendicular to the $\alpha$ direction, we will denote this segment as
$\alpha_2$, and the corresponding vector characterizing DM interaction will be denoted as $\vec{D}_{\alpha_2}$.
In our further analysis, only single component of each such DM vector is involved, namely for $\alpha=x,y,z$ we write $D^\alpha$ for $\alpha$-s component of $\vec{D}_{\alpha_2}$ and introduce $D=\sqrt{(D^x)^2+(D^y)^2+(D^z)^2}$.

Typical values  of various interaction constants for several "Kitaev materials"
were computed in~\cite{winter2016challenges}, they are  presented in the table below (in units of meV).

\begin{center}
	\begin{tabular}{ |c||c|c|c|c| } 
		\hline
		 &$K$ & $J$  & $\Gamma$ & $D$ \\ 
		 \hline \hline
		Na$_2$IrO$_3$ & $-16.2$ & $1.6$ & $2.1$ & $0.17$  \\ 
		$\alpha$-RuCl$_3$ & $-7.5$ & $-2.2$ & $8.0$ & $0.44$ \\
		$\alpha$-Li$_2$IrO$_3$ & $-13.0$ & $-4.6$ & $11.6$ & $0.44$ \\ 
		\hline
	\end{tabular}
\end{center}

For the general analysis of the role of perturbations,  the Hamiltonian of system can be presented in the form
\begin{eqnarray}
H=H_K+\sum_n V_n,
\end{eqnarray}
where $n$ enumerates different types of  local perturbations.
The spin-spin correlation function is  then given by the perturbation theory expansion
\begin{equation}
\label{spin-spin correlation function}
S^{\alpha \beta}_{ij}(t)= \langle T \sigma^\alpha_i(t)\sigma_j^\beta(0)\rangle =
\sum_n \frac{(-i)^n}{n!}\sum_{k_1 \ldots k_n}\int d\tau_1 \ldots d\tau_n\langle T \sigma^\alpha_i(t)\sigma_j^\beta(0) V_{k_1}(\tau_1)\ldots V_{k_n}(\tau_n)\rangle.
\end{equation}
The ground state average in Eq. (\ref{spin-spin correlation function}) vanishes if the product of operators
$\sigma^\alpha_i\sigma_j^\beta V_{k_1}\ldots V_{k_n}$  creates fluxes. To get a nonzero result,
the product of perturbation operators $V_{k_1}\ldots V_{k_n}$ should create the same fluxes as the product of
spin operators $\sigma^\alpha_i\sigma_j^\beta$ does. Two qualitatively different situations may occur, leading to either power-law, or 
exponential decay of spin correlations with a distance $|i-j|$.

\begin{figure}
	\includegraphics[scale=0.5]{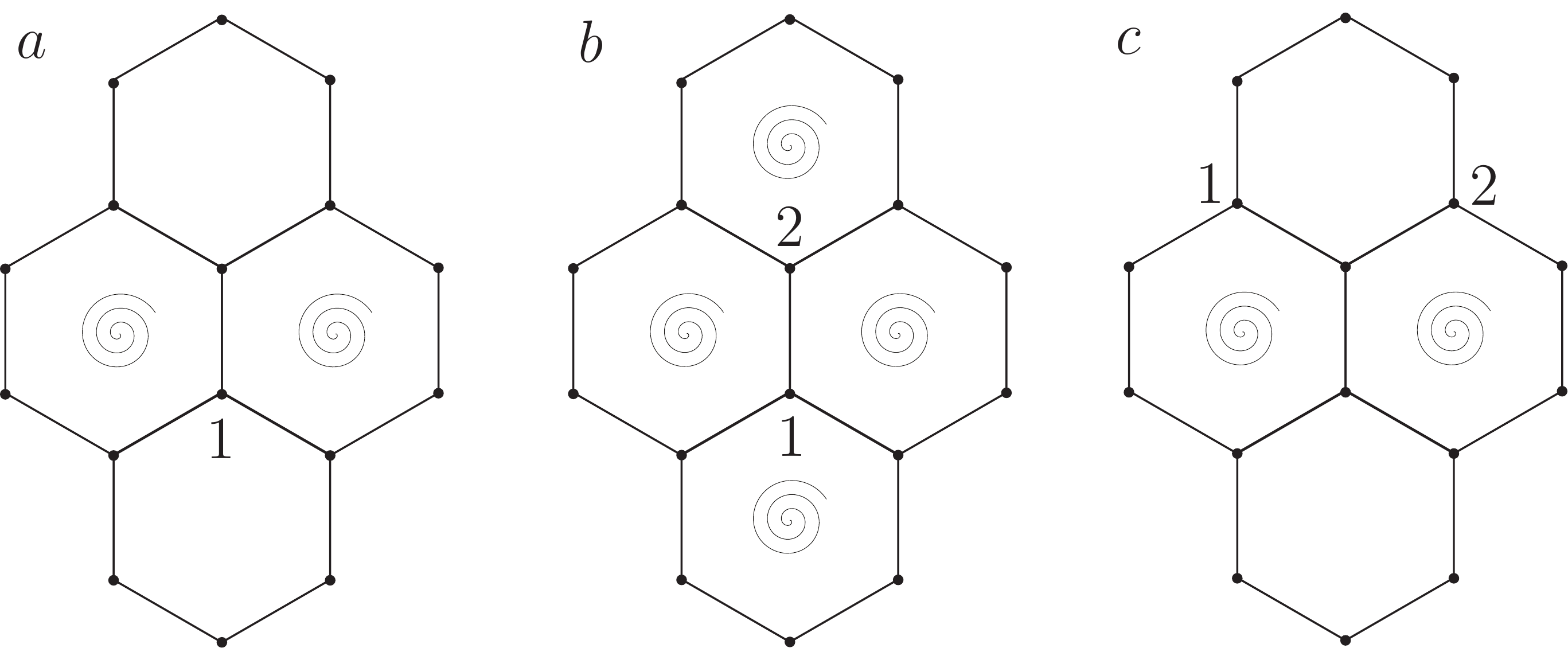}
	\caption{Flux patterns created by various perturbations. (2a) Magnetic field: $\sigma_1^z$, (2b) Heisenberg interaction: one specific term $\sigma_1^x\sigma_2^x$, (2c) DM interaction: one specific term $\sigma_1^y\sigma_2^x$.} 
\label{patterns}
\end{figure}

\textit{Power-law decay}  is realized if the  product of some number of perturbation operators $V_k$
can be represented in the form  $V_{k_1}\ldots V_{k_n}=V_{i,\alpha}V_{j,\beta}$, where
$V_{i,\alpha}$ creates the same fluxes as the spin operator $\sigma_i^\alpha$, and $V_{j,\beta}$ 
creates the same fluxes as the spin operator $\sigma_j^\beta$.  Then the necessary number $n$ of the operators 
$V_k$ in the product $V_{k_1}\ldots V_{k_n}$ does not grow with the distance $|i-j|$.
Such a situation is realized by the magnetic field perturbation \cite{TFK2011} with
\begin{eqnarray}
V_h^z=-h^z\sum_i \sigma_i^z,
\label{magnetic}
\end{eqnarray}
with flux pattern illustrated on the Fig. \ref{patterns}a as well as by the combined Heisenberg and pseudodipolar interactions~\cite{PhysRevLett.117.037209}
where $V=V_H+V_\Gamma$. The first mentioned case is the simplest one:
 long-range spin-spin correlation  $S^{zz}_{ij}(t)$  contains an overall coefficient
 $h_z^2$, as follows from the above analysis in terms of flux creation/annihilation.
The second case~\cite{PhysRevLett.117.037209} appears to be more delicate. Namely, a straightforward
counting of fluxes indicates appearance of long-range correlations with a coefficient
$\propto (\Gamma J)^2$.  Indeed, such terms appear in the perturbation theory, but they cancel completely due to some special symmetry reasons, and nonzero contribution was found in the next order of
perturbation theory, so it scales as $S^{\alpha \beta}_{ij}(t)  \propto (\Gamma J)^4$.
This observation demonstrates that analysis of fluxes is not sufficient, in general,
to determine the lowest necessary order of the perturbation that leads to long-range correlations.
More detailed analysis of the relevant symmetries will be presented below in the end of the Sec. \ref{sec_symm}.

\textit{Exponential decay of correlation functions with distance} takes place if  perturbation terms are not able
to annihilate  "locally" \, the fluxes created by spin operators $\sigma_i^\alpha$ and $\sigma_j^\beta$.
In such case the necessary result - complete flux annihilation - can be achieved in the 
$n$-th order of perturbation theory only, where $n\sim |i-j|$, thus in this case correlation functions decrease 
exponentially with a distance~\cite{mandal2011confinement}.  However, the above arguments do not imply
exponential decay of correlation functions with time $t$, thus in this situation unusual asymmetry of space-time correlations
may be expected; this issue needs further studies.
An example of exponential spatial decay of correlations is provided by purely Heisenberg perturbation
$\hat{V}_H$.  Each of such terms create fluxes in 4 plaquettes, as shown in Fig.2b. Consecutive action of
operators $\hat{V}_H$ ordered along the line between sites $i$ and $j$ consists in the moving pairs of fluxes
from the vicinity of the site $i$ where these fluxes were created, to the vicinity of the site $j$ where
they will be annihilated. Thus the minimal number of such operators is $\sim n$ and the whole
correlation function is bounded from above by $(J/K)^n \sim \exp\left(-n\ln\frac{K}{J}\right)$.

\section{Reduction of spin correlation function  to fermionic representation.}

Below we consider the Hamiltonians of the form  
\begin{equation}
\mathcal{H}=\mathcal{H_K}+\sum_{i\in even,\alpha} V_i^\alpha.
\label{hamiltonia_with_perturb_I}
\end{equation}
Here by $V_i^\alpha$  we denote  an operator which is\,  i) composed out of spin operators, and \, ii) creates the same
combination of fluxes as the operator $\sigma_i^\alpha$, or, equivalently, $\sigma^\alpha_{i+\alpha}$.
The site notation $i+\alpha$ means that this site is connected to the site $i$ by a bond of the $\alpha$ type.
Summation over $i\in even$ in the second term of (\ref{hamiltonia_with_perturb_I}) means that these $i$ sites
belong to the even sublattice of the honeycomb lattice; below we identify $V^\alpha_{i+\alpha}=V_i^\alpha$.
Particular examples of the models (\ref{hamiltonia_with_perturb_I}) are provided by
the Kitaev model with a magnetic (\ref{magnetic}) field and/or with the second-neighbours DM interaction (\ref{DM}). 

We are interested in the calculation of  the correlation function of spin operators $S^{\alpha\beta}_{ij}(t)=\langle T \sigma_i^\alpha(t) \sigma^\beta_j(0)\rangle$, for large  distances $r = |i-j| \gg 1$  between sites $i$ and $j$,
and at long time intervals  $Kt\gg1$.  Below we will show that such a correlation function can be represented
in terms of correlations of fermionic bilinear operators:
\begin{eqnarray}
S^{\alpha\beta}_{i,j}(t)=\langle T\sigma_i^\alpha(t)\sigma_j^\beta(0)\rangle=\langle T Q_i^\alpha(t) Q_j^\beta(0)\rangle,
\label{from_spins_to_fermions}
\end{eqnarray}
where 	
\begin{equation}
\label{Q_operator}
Q_i^\alpha=\frac{i}{2}\sum_{c_a,c_b\in \{c^{(i)}_1 \ldots c^{(i)}_n\}} V^\alpha_{i,ab} c_a c_b,
\end{equation}
and matrix elements:
\begin{equation}
V^\alpha_{i,ab}=\int^{\infty}_{0}d\tau   \langle   V_i^\alpha(\tau) \sigma_i^\alpha(0)c_a(0) c_b(0)\rangle
+\int^{0}_{-\infty} \langle   c_a(0)c_b(0) \sigma(0)_i^\alpha  V_i^\alpha(\tau)\rangle d\tau
\label{Q_operator_V}
 \end{equation}
In the above equation, the variables $c^{(i)}_a$ denote  fermionic operators
 located near $c_i$ and  defined in terms of the fermionic content of the operator product $\sigma_i^\alpha V_i^\alpha$:
\begin{eqnarray}
\sigma_i^\alpha V_i^\alpha=A_i c^{(i)}_1\ldots c^{(i)}_{n_i} U_i 
\label{sigmaV_through_majorana}
\end{eqnarray}
where $U_i$  is some product of  the bond integrals of motions ($u_{pq}$),
while $A_i$ is some constant. 

Matrix elements $V^\alpha_{i,ab}$ can also be written in terms of  spin correlation functions.
The simplest example is provided by $V_i^\alpha=\sigma_{i+\alpha}$. In this case
$Q^\alpha_{i}=i\gamma_i h^\alpha V_h c_{i} c_{i+\alpha} $,
where $\gamma_i=1$ if $i$ belongs to the even sublattice and $\gamma_i=-1$ otherwise. 
For the matrix element $V_h$ one finds then
$V_h=2 \int_0^\infty \langle \sigma_i^z(\tau) \sigma_i^z(0)\rangle d\tau$. 

Now we proceed with the derivation of the representation (\ref{from_spins_to_fermions}), (\ref{Q_operator}) and (\ref{Q_operator_V})
for spin correlation function. If $V_i^\alpha$ is a product of spin operators, the first non-zero term of the perturbation expansion  (\ref{spin-spin correlation function}) is of the second order in $V_i^\alpha$:
 \begin{equation}
S^{\alpha \beta}_{ij}(t)=\langle T\sigma^\alpha_i(t) \sigma^\beta_j(0)\rangle=-\int d\tau_1 d\tau_2 \langle T\sigma^\alpha_i(t) \sigma^\beta_j(0) V^\alpha_i(\tau_1) V^\beta_j(\tau_2)\rangle.
\label{Int12}
\end{equation}
 The main contribution to the integrals in Eq.(\ref{Int12}) comes from the region where $|t-\tau_1|K\sim1$ and 
$|\tau_2|K\sim1$, see Ref. \onlinecite{TFK2011}.  There are 4 possible variants of the time  ordering in this region. 
Let us consider just  one of them as an example:  
\begin{equation}
S^{(I)}=-\int^{t}_{0 } d\tau_1 \int^{\tau_1}_{0}d\tau_2 \langle \sigma^\alpha_i(t) V^\alpha_i(\tau_1) V^\beta_j(\tau_2) \sigma^\beta_j(0) \rangle.
\label{cor_function_contribution}
\end{equation}
To shorten further notations, we define operators $\tilde{V}_i^\alpha$ via the following identity:
\begin{eqnarray}
V_i^\alpha\tilde{V}_i^\alpha=[H,V_i^\alpha].
\label{commutation_relations}
\end{eqnarray}
Operators with tilde sign  do not create fluxes, they are of the second order in  Majorana variables, 
and each of the fermionic bilinear contains fermions which belong to different  sublattices.
With the definition (\ref{commutation_relations}), we represent correlation function
(\ref{cor_function_contribution}) in the form
\begin{equation}
S^{(I)}=-\int^{t}_{0 } d\tau_1 \int^{\tau_1}_{0}d\tau_2 \langle e^{iHt} \sigma^\alpha_i V^\alpha_i e^{-i(H+\tilde{V}_i^\alpha)(t-\tau_1)}e^{-iH(\tau_1-\tau_2)}e^{-i(H+\tilde{V}^\beta_j)\tau_2} V^\beta_j\sigma^\beta_j \rangle,
\label{tilda}
\end{equation} 
which can be further transformed using representation of operators entering  Eq.(\ref{tilda})
 in terms of $c_a$ and $u_{ab}$ via Eq.(\ref{sigmaV_through_majorana}). The result of this transformation
reads:
\begin{equation}
S^{(I)}=-\int^{t}_{0 } d\tau_1 \int^{\tau_1}_{0}d\tau_2 \langle T A_i c^{(i)}_{1}(t)\ldots c^{(i)}_{n_i}(t) A_j c^{(j)}_{1}(0)\ldots c^{(j)}_{n_j}(0)e^{-i(\int_{\tau_1}^{t}\tilde{V}_i^\alpha(\tau^\prime)d\tau^\prime+\int^{\tau_2}_0 \tilde{V}_j^\beta(\tau^\prime)d\tau^\prime)}\rangle.
\label{cor_function_contribution_2}
\end{equation} 

The above correlation function  describes non-interacting  Majorana fermions living  in the time-dependent potential, which is switched on/off  for some intervals  of time.  
Thus we come to  a kind of problem similar to the celebrated Fermi-edge singularities~\cite{nozieres1969singularities}. 
The key difference consists in the zero density of states at the Fermi level in our problem (due to conical
spectrum and fixed  at $\epsilon=0$ position of the  Fermi-level).
This is the reason for the absence of the "orthogonality catastrophe" \, ~\cite{anderson1967infrared} in our case.

\begin{figure}
\includegraphics[scale=0.6]{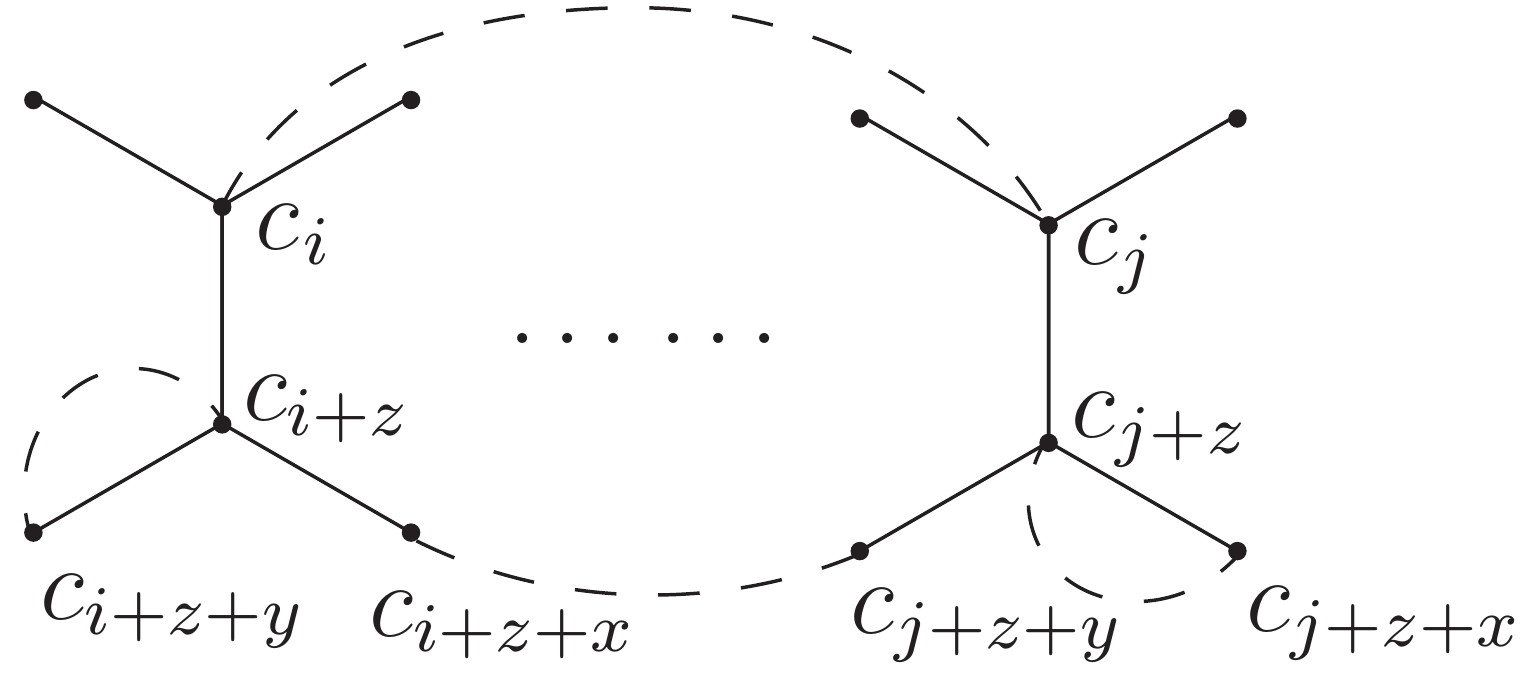}
\caption{One of the diagram for evaluation of the spin correlation function in the presence DM interaction: dashed lines are for Majorana Green function. Two of the pairings involve fermions, separated by large distance with the rest paired in the vicinity of each of the spin operators (applied at points $i$ and $j$).}
\label{diag_for_cor_func}
\end{figure}

We analyse now the average of a product of  Majorana operators,  Eq. (\ref{cor_function_contribution_2}), by means
of diagram technique in terms of free Majorana Green functions $G_{ab}(t)=\langle T c_a(t) c_b(0)\rangle$, using
Wick theorem. The key observation is based upon the fact that we are studing infrared asymptotics, $(r, Kt) \gg 1$,
and the corresponding Green functions decay  rather quickly with distance and time,
 $G_{ij}\propto (\max(r,Kt)^{-2})$,
 where $r$ is distance between $i$ and $j$. For this reason, any irreducible diagram, entering the average in
 (\ref{cor_function_contribution_2}), should contain exactly two  "nonlocal" \, Green functions
(nonlocal is an average of 2 fermion operators such that one of them is located near $c_i$ and the other near $c_j$). 
Apart from these two nonlocal lines, an arbitrary number of local Majorana Green functions (containing
both fermionic operators located near point $i$ or near point $j$)  may enter various
diagrams originating from Eq.(\ref{cor_function_contribution_2}).  One of typical diagrams of such a type
is shown in Fig.~\ref{diag_for_cor_func} for the specific case of the DM perturbation.

Summation of all these diagrams can be represented analytically in the following form:
\begin{eqnarray}
S^{(I)}=-\frac{1}{4}\sum_{a,b,p,q}\langle T c_a(t)c_b(t)c_p(0)c_q(0)\rangle\int^{t}_{0 } d\tau_1 \int^{\tau_1}_{0}d\tau_2 \langle  c_a(t)c_b(t) \sigma(t)_i^\alpha  V_i^\alpha(\tau_1)\rangle  \langle    V_j^\beta(\tau_2) \sigma_j^\beta(0)c_p(0) c_q(0)\rangle.
\label{last_expression_for_contribution}
\end{eqnarray}
In the expression above, the first average of four Majorana operators contains two non-local Green functions. 
Second and third averages contains  two  sets of local diagrams, located near the points $i$ and $j$ respectively.
Identity $c_a^2=1$ was used to represent the diagrams in the form (\ref{last_expression_for_contribution}). 
The main contribution to  the integral over $d\tau_1$ and $d\tau_2$  in  Eq.(\ref{last_expression_for_contribution})
comes from  the region  where $K|t-\tau_1|\sim1$ and $K\tau_2\sim1$;  taking into account the condition
$K|t|\gg1$, we can extend the limits of integration over $d\tau_{1,2}$ to infinity.
Finally, taking into account all possible versions of time ordering, we obtain 
representaion (\ref{from_spins_to_fermions}-\ref{Q_operator_V}) for  spin correlation function.

The above analysis can be generalized to the case when perturbation $V$ contain several operators $V_1$, $\ldots$, $V_n$,
and their product creates the same pattern of fluxes as $\sigma_i^\alpha$ does, as well as
for the operator  $\sigma_j^\beta$.  We still can write spin-spin correlation function in the form of 
 Eq. ( \ref{from_spins_to_fermions}), while expressions (\ref{Q_operator},\ref{Q_operator_V}) should be modified.
Namely, Majorana operators $c_a$ and $c_b$ which enter $Q_i^\alpha$ are those operators, which are contained
in the operator product $\sigma_i^\alpha V_1 \ldots V_n$  when it is expressed in terms of Maiorana variables
 $c_i$ and bond variables $u_{ij}$. Generalized expression for matrix elements  $V_{ab}$ now reads as follows:
\begin{eqnarray}
\label{Vab}
V_{ab}=\sum_{k=0}^{n} \sum_{\zeta=\langle i_1 \ldots i_n \rangle\in S_n} \xi_{ab,\zeta,k}\int_0^\infty d\tau_1 \int_0^{\tau_1}d\tau_2\ldots \int_0^{\tau_{k-1}}d\tau_k \int_{-\infty}^0 d\tau_{k+1} \ldots \int_{-\infty}^{\tau_{n-1}} d\tau_n \nonumber \\
\langle V_{i_1}(\tau_1) \ldots V_{i_k}(\tau_k) c_a(0)c_b(0)\sigma_i^\alpha(0) V_{i_{k+1}}(\tau_{k+1})\ldots V_{i_n}(\tau_n)\rangle.
\label{quadratic_correction_general}
\end{eqnarray}
Here $\zeta$ is some permutation of numbers  $1,\ldots n$,  and $\xi_{ab,\zeta,k}=\pm1$ are defined via 
an identity  $c_ac_b V_{i_1}\ldots V_{i_k}=\xi_{ab,\zeta,k}  V_{i_1}\ldots V_{i_k} c_ac_b $.
 In the term with  $k=0$ all $\tau_i<0$, while  the  term with  $k=n$ contains all $\tau_i>0$.

Let us now see how this general treatment applies to some particular example. We will consider Kitaev model, perturbed by small magnetic field, studied in Ref. \onlinecite{TFK2011}. This is described by Eq. (\ref{hamiltonia_with_perturb_I}) with $V_i^\alpha=-h^\alpha(\sigma_i^\alpha+\sigma_{i+\alpha}^\alpha)$, giving 
\begin{eqnarray}
\label{Q}
Q^\alpha_{i}=i\gamma_i h^\alpha V_h c_{i} c_{i+\alpha},
\end{eqnarray}
with $\gamma_i=\pm 1$ for lattice site $i$ in the even/odd sublattice correspondingly and numerical coefficient $V_h=2 \int_0^\infty \langle \sigma_i^z(\tau) \sigma_i^z(0)\rangle d\tau$. Substituting Eq. (\ref{Q}) to Eq. (\ref{from_spins_to_fermions}), we obtain at large $\max(r,t)$ (but far from the mass shell $|r-\sqrt{3}Kt|\gg1$):
\begin{eqnarray}
S^{\alpha\beta}(\mathbf{r},t)_{ij}= \frac{h^\alpha h^\beta V_h^2}{\pi^2 (r^2-3K^2t^2)^3}\{r^2\sin\left[\frac{2\pi}{3}(\mathbf{e}_x,\mathbf{n}_\beta+\mathbf{r})-\phi_x\right]\sin\left[\frac{2\pi}{3}(\mathbf{e}_x,\mathbf{n}_\alpha-\mathbf{r})-\phi_x\right] \nonumber\\-3K^2t^2\cos\left[\frac{2\pi}{3}(\mathbf{e}_x,(\mathbf{n}_\alpha-\mathbf{n}_\beta-\mathbf{r}))\right]\cos\left[\frac{2\pi}{3}(\mathbf{r},\mathbf{e}_x)\right]\}.
\label{S1ab}
\end{eqnarray}
In this equation, we have assumed for simplicity that $i$ and $j$ belong to the even sublattice and introduced $\mathbf{n}_x=2\mathbf{n}_1$, $\mathbf{n}_y=2\mathbf{n}_2$, $\mathbf{n}_z=\mathbf{0}$, $\mathbf{e}_x$ for the unit vector along $x$ axis and $\phi_x$ for the angle between $\mathbf{r}$ and  $\mathbf{e}_x$. In the case of either
$i$ or $j$ belonging to the odd sublattice, one  can use identity $Q_{i+\alpha}^\alpha=Q_i^\alpha$ in order to
generalize Eq.(\ref{S1ab}).

The structure factor which is Fourier transform of correlation function in this case demonstrates interesting behaviour near three points in the Brillouin zone, located at $(0,0)$, $\mathbf{ q}_+=2\mathbf{e_x}$ and $\mathbf{q}_-=-\mathbf{q}_+$. 
Near $(0,0)$ it is given by
\begin{eqnarray}
S^{\alpha\beta}(\mathbf{q},\omega)-S(\mathbf{0},0)= \frac{h^\alpha h^\beta V_h^2\theta(\omega^2-3K^2q^2)}{8\sqrt{\omega^2-3K^2q^2}}\{q^2+(2\omega^2-3K^2q^2)\cos(\frac{2\pi}{3}(\mathbf{e}_x,\mathbf{n}_\alpha-\mathbf{n}_\beta))\},
\end{eqnarray}
while near $\mathbf{q}_{\pm}$ the structure factor varies as
\begin{eqnarray}
S^{\alpha\beta}(\mathbf{q},\omega)-S(\mathbf{q}_\pm,0)= \frac{h^\alpha h^\beta V_h^2\theta(\omega^2-3K^2q^2)}{4}\sqrt{\omega^2-3K^2q^2}e^{\mp i\frac{2\pi}{3}(\mathbf{e}_x,\mathbf{n}_\alpha-\mathbf{n}_\beta)}.
\end{eqnarray}


\section{Effective Majorana Hamiltonian}
As we have discussed above, one of the effects of the perturbations on the Kitaev model is to couple the spin density to local bilinear Fermionic operators. However, it is natural to expect that a general perturbation
is also able to modify the dynamics of Fermionic operators themselves. Indeed, as was demonstrated by Kitaev~\cite{Kitaev06}, magnetic field $h$ treated perturbatively in the 3rd order opens a gap in the fermionic spectrum, proportional to $h_xh_yh_z$. We will now analyse the effects of high-order terms in the perturbation theory for a more generic perturbation. Consider the following Hamiltonian:
\begin{eqnarray}
H=H_K+\sum_n\left[V_{n,1}+V_{n,2}\right].
\label{Hamiltonia_for_gap}
\end{eqnarray}
In this equation, we have explicitly introduced two contributions to the perturbing Hamiltonian, $V_{n,1}$ and $V_{n,2}$ which create the same flux pattern when applied to the ground state of unperturbed model.
We will assume that $V_{n,\xi}$ are products of spin operators. Our goal will be to study the perturbation theory series for the Fermionic Green function and resum it in the low-energy limit. Let us first present our result. We have found that corrections to the Green function in this limit are described
by the following correction to the Fermionic Hamiltonian:
\begin{equation}
\delta H=\frac{(-i)}{4}\sum_{m} \sum_{p,q} c_{p}c_{q}V_{m,pq},
\label{quadratic_correction_3}
\end{equation}
with numerical coefficients given by:
\begin{equation}
V_{m,pq}=\int_0^\infty d\tau \langle V_{m,1}(\tau) V_{m,2}(0)c_p(0) c_q(0) \rangle +\int_{-\infty}^0 \langle c_p(0)c_q(0) V_{m,2}(0) V_{m,1}(\tau)\rangle d\tau +(1\leftrightarrow 2).
\label{Vmpq}
\end{equation}
Majorana fermions $c_p$ and $c_q$ entering Eq.(\ref{Vmpq})
are contained in the operator product $V_{m,1}V_{m,2}$, in the same sense as it was indicated in 
Eq. (\ref{sigmaV_through_majorana})  regarding operator product  $ \sigma_i^\alpha V_i^\alpha$.

Let us explain how this result can be derived. We start from perturbation theory series for the Green function:
\begin{equation}
G_{ij}(t)=\sum_{n}\frac{(-i)^n}{n!}\sum_{m_i,\xi_i,\zeta_i} \int d\tau_1 \ldots  d\tau_n \langle T c_i(t) c_j(0)V_{m_1,\xi_1}(\tau_1)\ldots V_{m_n,\xi_n}(\tau_n)\rangle
\label{Green_begin}
\end{equation}
The average of an operator in Eq. (\ref{Green_begin}) is non-zero only when it is flux-free, which requires $n$ to be even and all $V_{m,\xi}$ to be separable into flux-free pairs. As operators in each pair appear at different instants of time, the flux does exist for short (of the order of $K^{-1}$) periods of time.  As $tK\gg1$, time intervals between pairs are larger than $K^{-1}$. Thus, we can rewrite (\ref{Green_begin}):
\begin{eqnarray}
G_{ij}(t)=\sum_{n}\frac{(-i)^{2n}}{n!2^n}\sum_{m_i,\xi_i,\zeta_i} \int d\tau_1 \ldots  d\tau_{2m} \langle T c_i(t) c_j(0)V_{m_1,\xi_1}(\tau_1)V_{m_1,\zeta_1}(\tau_2)\ldots V_{m_n,\xi_n}(\tau_{2n-1})V_{m_n,\zeta_n}(\tau_{2n})\rangle \nonumber \\
\label{Green_begin_1}
\end{eqnarray}
For each $n$ there are $2^n$ different contributions ($\tau_{2l-1}>\tau_{2l}$ or $\tau_{2l}>\tau_{2l-1}$ for all $l=1\ldots n$). Let us consider one of them:
\begin{eqnarray}
\delta G_{ij}(t)=\sum_{n}\frac{(-i)^{2n}}{2^n}\sum_{m_i,\xi_i,\zeta_i} \int_{-\infty}^\infty d\tau_1 \int_{-\infty}^{\tau_1}d\tau_2 \ldots  \int_{-\infty}^{\tau_{2m}-1}d\tau_{2m} \nonumber \\ \langle T c_i(t) c_j(0)V_{m_1,\xi_1}(\tau_1)V_{m_1,\zeta_1}(\tau_2)\ldots V_{m_n,\xi_n}(\tau_{2n-1})V_{m_n,\zeta_n}(\tau_{2n})\rangle \nonumber. \\
\label{Green_begin_2}
\end{eqnarray}
Using commutation relations (\ref{commutation_relations}), we can now bring together operators in each flux-free pair together (similarly to the calculation of the spin-spin correlation function):
\begin{eqnarray}
\delta G_{ij}(t)=\sum_{n}\frac{(-i)^{2n}}{2^n}\sum_{m_i,\xi_i,\zeta_i} \int_{-\infty}^\infty d\tau_1 \int_{-\infty}^{\tau_1}d\tau_2 \ldots  \int_{-\infty}^{\tau_{2m}-1}d\tau_{2m} \nonumber \\ \langle T c_i(t) c_j(0)V_{m_1,\xi_1}(\tau_1)V_{m_1,\zeta_1}(\tau_1)\ldots V_{m_n,\xi_n}(\tau_{2n-1})V_{m_n,\zeta_n}(\tau_{2n-1})e^{-i\int \tilde{V}(\tau^\prime)d\tau^\prime}\rangle, \nonumber \\
\label{Green_begin_3}
\end{eqnarray}
where $\tilde{V}$ stays for the scattering potential induced by the fluxes. Similarly to Eq. (\ref{cor_function_contribution_2}), to obtain the main contribution we need only an open line contributions containing non-local Green functions, joining distant points, with all other Green functions being are 'local'. Upon this selection, we obtain the following expression: 
\begin{eqnarray}
\delta G_{ij}(t)=\sum_{n}\frac{(-i)^{2n}}{2^{2n}}\sum_{m_1,\xi_1,\zeta_1 \ldots m_n, \xi_n,\zeta_n}\sum_{p_1,q_1,\ldots p_n,q_n} \int_{-\infty}^\infty d\tau_1 \int_{-\infty}^{\tau_1}d\tau_2 \ldots  \int_{-\infty}^{\tau_{2m}-1}d\tau_{2m} \nonumber \\ \langle T  c_i(t) c_j(0) c_{p_1}(\tau_1)c_{p_2}(\tau_1)\ldots c_{p_n}(\tau_n)c_{q_n}(\tau_n)\rangle\nonumber \\ \langle c_{p_1}(\tau_1)c_{q_1}(\tau_1)V_{m_1,\xi_1}(\tau_1)V_{m_1,\zeta_1}(\tau_2)\rangle\ldots \rangle c_{p_n}(\tau_{2n-1})c_{q_n}(\tau_{2n-1})V_{m_n,\xi_n}(\tau_{2n-1})V_{m_n,\zeta_n}(\tau_{2n})\rangle \nonumber. \\
\label{Green_begin_4}
\end{eqnarray}
As in Eq. (\ref{last_expression_for_contribution}), the first average in Eq. (\ref{Green_begin_4}) describes non-local diagram pairings and the remaining $n$ averages describe  local Green function near $m_1$,$\ldots$,$m_n$ respectively. Indices $p_i$ and $q_i$ enumerate all Majoranas which appear in the expression for $V_{m_i,1}V_{m_i,2}$ expanded using Eq. (\ref{sigmaV_through_majorana}). After summation of all $2^n$ contributions (a number of different time orderings), we find that correction to the Green function is reproduced by effective Hamiltonian in Eq. (\ref{quadratic_correction_3}).

 As a similar result for the correlation function, see Eq. (\ref{quadratic_correction_general}), this result for effective Hamiltonian can be generalized. In a more general situation, one may have not a pair of perturbing terms $V_1,\;V_2$, which is flux-free but a larger combination, say, $V_1,\;\ldots,\;V_n$. In this case, correction to the low-energy Hamiltonian becomes as follows: 
\begin{equation}
\delta H=\frac{-i}{4}\sum_{a,b} V_{ab}c_ac_b,
\end{equation} where $c_a$ and $c_b$ are fermions, which enter the product $V_1\ldots V_n$, expanded in terms of Majoranas $c_i$ and conserved quantities $u_{ij}$. The matrix elements in this effective Hamiltonian read: 
\begin{eqnarray}
V_{ab}=\sum_{k=1}^{n}\sum_{\zeta=\langle i_1 \ldots i_n\rangle \in S_n} \xi_{ab,\zeta,k} \int_0^\infty d\tau_1 \int_0^{\tau_1}d\tau_2 \ldots \int_0^{\tau_{k-1}} d\tau_k \int_{-\infty}^0 d\tau_{k+2} \ldots \int_{-\infty}^{\tau_{n-1}} d\tau_n \nonumber \\ 
\langle  V_{i_1}(\tau_1) \ldots V_{i_k}(\tau_k) c_a(0) c_b(0) V_{i_{k+1}}(0) V_{i_{k+2}}(\tau_{k+2})\ldots V_{i_n}(\tau_n)\rangle.
\label{coefficent_in_Hamiltonian}
\end{eqnarray}   
In this equation, $\zeta$ is a permutation of $1,\ldots n$ and $\xi_{ab,\zeta,k}=\pm1$ is introduced according to the following identity:
\begin{equation}
c_ac_b V_{i_1}\ldots V_{i_k}=\xi_{ab,\zeta,k}  V_{i_1}\ldots V_{i_k} c_ac_b.
\end{equation}
The boundary cases of $\tau_i$ in Eq. (\ref{coefficent_in_Hamiltonian}) should be understood as follows: for $k=0$ all $\tau_i<0$ and for $k=n$ all $\tau_i>0$.


\section{Implications of time-reversal symmetry}
\label{sec_symm}
In general, evaluation of matrix elements $V_{ab}$ in (\ref{Q_operator_V}) and (\ref{quadratic_correction_3})  is a complicated task. However, some of their properties follow directly from the properties of the spin perturbation with respect to time inversion. Expressions for matrix elements have similar form: 
\begin{eqnarray}
V=\int_0^\infty \langle V_1(\tau) V_2(0)\rangle d\tau -\int_{-\infty}^0 \langle V_2^\dagger(0) V^\dagger_1(\tau)\rangle d\tau,
\label{symmetry} 
\end{eqnarray} 
where $V_1=V_i^\alpha$ and $V_2=\sigma_i^\alpha c_a c_b$ (note that $V_{1,2}$ are not necessarily Hermitian). These operators can be written in terms of spins operators with the same flux patterns. The coefficients $V_{a,b}$ in Eqs. (\ref{Q_operator}), (\ref{quadratic_correction_3}) have the same structure. It is convenient to represent operators $V_{1,2}$ as $V_\alpha=\eta_\alpha C_\alpha B_\alpha$ with $\alpha=1,2$, $\eta$ for some constant, $C$ for a product of $c_i$ and $B$ for a product of $b_i$. Using commutation relations (\ref{commutation_relations}) we can write Eq. (\ref{symmetry}) (up to the sign of the whole expression) as follows:
\begin{eqnarray}
\pm V=\eta_1 \eta_2 \int_0^\infty \langle T C_1(\tau) C_2(0)B_1(0)B_2(0)e^{-i\int_0^\tau \tilde{B}_1(\tau^\prime)d\tau^\prime}\rangle-\eta_1^*\eta_2^*\int_{-\infty}^0 \langle T B_2^\dagger(0)B_1^\dagger(0) C_2^\dagger(0)  C_1^\dagger(\tau) e^{-i\int_\tau^0 \tilde{B}_1(\tau^\prime)d\tau^\prime}\rangle d\tau \nonumber. \\
\label{CC}
\end{eqnarray}  
The operator $\tilde{B}_1$ is defined similarly to the operator $\tilde{V}$, see Eq. (\ref{commutation_relations}). The operator $B_1B_2$ contains only $b_i$, it is flux-free and equals to some constant $b$ under the average. In what follows, we will consider both averages in Eq. (\ref{CC}) expanding it via Wick theorem. As $c_i=c_i^\dagger$, all operators in different terms in (\ref{CC}) are the same but are ordered differently. Importantly, $G_{11}$ and $G_{22}$ are odd functions of time and $G_{12}$ and $G_{21}$ are even.  To rearrange $C_2^\dagger C_1^\dagger$ into $C_1C_2$ under the sign of  T-average  we need $\frac{N(N-1)}{2}$ transpositions, where $N$ is the total number of fermionic operators in $C_1C_2$. We can change a sign of integration time variables in the second term and use properties of time reversal symmetry of Green functions. After such a change, it will have acquire a sign $(-1)^m$ where $m$ is a number of Green functions $G_{11}$ and $G_{22}$. Parity of $m$ is equal for each terms. This follows from the properties of $\tilde{B}$, which is quadratic, with each term being a product of fermions from different sublattices. So we can write:
\begin{equation}
\pm V=(b \eta_1 \eta_2-(-1)^{\frac{N(N-1)}{2}+m} b^\dagger \eta_1^*\eta_2^*) \int_0^\infty \langle T C_1(\tau) C_2(0)e^{-i\int_0^\tau \tilde{B}_1(\tau^\prime)d\tau^\prime}\rangle.
\label{difference_v}
\end{equation}    
Now, we note that $m\equiv(\frac{N}{2}-N_2) (\mod~2)$, where $N_2$ is a number of fermions from second sublattice in the operator $C_1C_2$. As $N$ is even, $\frac{N(N-1)}{2}\equiv \frac{N}{2}(\mod~2)$ so $m+\frac{N(N-1)}{2}\equiv -N_2 (\mod~2)$. The last step is to use time-reversal  symmetry:
\begin{eqnarray}
 Tc_{i,1}T^{-1}=c_{i,1},\; Tc_{i,2}T^{-1}=-c_{i,2},
 \label{time_reverse_for_c}
\end{eqnarray} so that:
\begin{eqnarray}
b^\dagger\eta_1^*\eta_2^*V_1V_2=(-1)^{N_2}b\eta_1\eta_2TV_1V_2T^{-1}.
\end{eqnarray}   
On the other hand, as $V_1$ and $V_2$ are spin-operators, $TV_1V_2T^{-1}=(-1)^\zeta V_1V_2$ where $\zeta=0,1$ so we can rewrite (\ref{difference_v}):
\begin{eqnarray}
\pm V=(b \eta_1 \eta_2)(1-(-1)^\zeta) \int_0^\infty \langle T C_1(\tau) C_2(0)B_1(0)B_2(0)e^{-i\int_0^\tau \tilde{B}_1(\tau^\prime)d\tau^\prime}\rangle.
\end{eqnarray}
As a result, if $\zeta=0$ than $V=0$ and if $\zeta=1$, $V\ne0$. Let us now apply our result to analysis of expressions for effective spin operator, Eq. (\ref{Q_operator}) and correction to the Hamiltonian, Eq. (\ref{quadratic_correction_3}) arising due to perturbations.

In Eq. (\ref{Q_operator}), one has $V_1=V_i^\alpha$ and $V_2=\sigma_i^\alpha c_a c_b$. Hence, $V_{ab}$ does not vanish if $TV_i^\alpha\sigma_i^\alpha c_a c_bT^{-1}=-V_i^\alpha\sigma_i^\alpha c_a c_b$. If $TV_i^\alpha T^{-1}=V_i^\alpha$ than $c_a$ and $c_b$ belong to the same sublattice, according to Eq. (\ref{time_reverse_for_c}). Contrarily, if $TV_i^\alpha T^{-1}=-V_i^\alpha$, they belong to different sublattices. Note that if in both  $Q^i_\alpha$ and $Q_i^\beta$ there is a term$\propto c_{a,1}c_{b,2}$, the correlation function oscillates at large r \cite{TFK2011}, in the opposite case the oscillations are absent \cite{PhysRevLett.117.037209}. 

This analysis helps to understand why spin-spin correlation function in Kitaev model with Heisenberg and pseudodipolar interaction is non-zero only in eighth order of the perturbation theory\cite{PhysRevLett.117.037209}, while according to the flux counting, the first non-zero appears already in the fourth order. Let us show that $4$th order contribution actually vanishes. In order to do so, we will make use of Eqs (\ref{Q_operator}) and (\ref{Vab}). Heisenberg and pseudodipolar interactions are both time-reversal invariant. According to our symmetry analysis, expression for $Q$ can be composed only by products of Majorana fermions from the same sublattice.  At the same time, these Fermionic operators should appear from expansion of the product of a spin operator and perturbation operators in terms of Majoranas. However, it is easy to check that in this expansion there are only two Fermions which belong to different sublattices. Hence, this perturbation can not contribute to coupling between Majoranas and spin operators.

We can also apply this reasoning to effective Hamiltonian: if in Eq. (\ref{quadratic_correction_3}) one has $TV_{m,\xi}V_{m,\zeta}T^{-1}=V_{m,\xi}V_{m,\zeta}$ than  $c_p$ and $c_q$ are from different sublattices. According to Eq. (\ref{free_hamiltonian}), effect of such a correction is to change the position of the conic point. If  $TV_{m,\xi}V_{m,\zeta}T^{-1}=-V_{m,\xi}V_{m,\zeta}$ than $c_p$ and $c_q$ belong to different sublattices and the spectrum of Majorana Fermions becomes gapped.


\section{Kitaev model in the presence of DM interaction: spin-spin correlations}

Let us now discuss the applications of the ideas outlined above to Kitaev model, perturbed by DM interaction.
The specifics of DM interaction, see Eq. (\ref{DM}) is that it has a term with a flux pattern like a spin operator, see Fig. \ref{patterns}c. As a result, 
it can potentially lead to a power-law decay of the spin-spin correlation function. We need to check, however, whether it is indeed the case, as naive 
flux counting may not always be enough (see discussion in the end of the section \ref{sec_symm}). In this section, we calculate the spin correlation function and spin structure factor in the presence of DM interaction. We also argue that magnetic field together with DM interaction opens a gap in the Majorana spectrum already in the first order in the magnetic field.

To calculate the spin correlation function we need to find local in space operator $Q$, defined in Eq. (\ref{Q_operator}). Let us start with the case of $\alpha=z$. There are two contributions in Eq. (\ref{DM}) which create spin-like fluxes: $D^z \sigma^x_{i+x}\sigma_{i+y}^y$ and $D^z \sigma_{i+z+x}^x\sigma_{i+z+y}^y$, where $i+x$ denotes a position of lattice site, connected to $i$ by $x$-edge. 
The definition of the component $D^z$, as well as  $D^x$ and $D^y$, was provided
right below the table, located after Eq.(\ref{DM}). As a result, operator $Q$ can be written as follows:
\begin{eqnarray}
Q_i^z=iD^z( A c_{i+x} c_{i+y}+B c_i c_{i+z+x}-B c_i c_{i+z+y}+C c_{i+z+x }c_{i+z+y}),
\label{Q_z_DM}
\end{eqnarray}    
where $A,\;B,\;C$ are some numerical constants. 
The rotational symmetry allows to write to $Q_i^x$ and $Q_i^y$ in the following form:
\begin{eqnarray}
Q_i^x=iD^x(Ac_{i+y}c_{i+z}+Bc_i c_{i+x+y}-Bc_ic_{i+x+z}+Cc_{i+x+y}c_{i+x+z}) \\
Q_i^y=iD^y(Ac_{i+z}c_{i+x}+Bc_i c_{i+y+z}-Bc_ic_{i+y+x}+Cc_{i+y+z}c_{i+y+x})
\end{eqnarray} 
These formulae can be rewritten compactly using notation $\alpha\beta(\gamma)$ 
(introduced below Eq. (\ref{Kitaev_Hamiltonian})):
\begin{eqnarray}
Q_i^\gamma=Q_i^{\alpha\beta(\gamma)}=iD^\gamma(Ac_{i+\alpha}c_{i+\beta}+Bc_i c_{i+\gamma+\alpha}-Bc_ic_{i+\gamma+\beta}+Cc_{i+\gamma+\alpha}c_{i+\gamma+\beta})
\end{eqnarray}
This brings us to  the correlation function:
\begin{eqnarray}
S^{\alpha\beta}_{a,b}(\mathbf{r},t)=\frac{3 D^\alpha D^\beta }{4\pi^2(r^2-3K^2t^2)^3} \left[((A-C)^2+4B^2)\hat\sigma_0+4B(A-C)\hat\sigma_1\right](\hat\sigma_1 r^2-\hat\sigma_0 3K^2t^2).
\end{eqnarray}
In this equation, $a,\;b$ enumerate sublattices and $\hat\sigma_0$ is the identity matrix $\hat\sigma_1$ is the first Pauli's matrix in space $a,b$.
Structure factor of this model reads:
\begin{eqnarray}
S^{\alpha\beta}_{a,b}(\mathbf{q},\omega)=\frac{3 D^\alpha D^\beta \theta(\omega^2-3K^2q^2)}{16\sqrt{\omega^2-3K^2q^2}}  \left[((A-C)^2+4B^2)\hat\sigma_0+4B(A-C)\hat\sigma_1\right](\hat\sigma_1 (2\omega^2-9K^2q^2)+\hat\sigma_0(2\omega^2-3K^2q^2)) \nonumber \\
\end{eqnarray}
 
The spin correlation function  above was derived in the lowest order of perturbation theory. This is a reasonable approximation, as DM interaction, treated in higher orders, does not generate a gap in the spectrum of low-lying excitations. Nevertheless, the presence of DM interaction modifies drastically response of the system to magnetic field $\mathbf{h}$, which becomes able to produce a gap for the low-lying excitations already in the first order in $|\mathbf{h}|$. In fact, quadratic correction to Hamiltonian in this case reads:
\begin{eqnarray}
Q_{hD}=\frac{i}{2} \sum_{i, \alpha\beta(\gamma)} h^\gamma D^\gamma(\tilde{A}c_{i+\alpha}c_{i+\beta}+\tilde{B}c_i c_{i+\gamma+\alpha}-\tilde{B}c_ic_{i+\gamma+\beta}+\tilde{C}c_{i+\gamma+\alpha}c_{i+\gamma+\beta})=\nonumber\\ =\frac{i}{2}\sum_{i, \alpha\beta(\gamma)} (h^\gamma D^\gamma(\tilde{A}+\tilde{C})+\tilde{B}(h^\alpha D^\alpha+h^\beta D^\beta))c_{i+\alpha}c_{i+\beta}=\frac{i}{2}\sum \Gamma_{\alpha\beta}c_{i+\alpha}c_{i+\beta},
\end{eqnarray}
where the terms which modify the position of the conic points and shape of the spectrum are omitted and $\tilde{A},\tilde{B},\tilde{C}$ are numerical constants.

In order to evaluate the effect on the spectrum, consider this correction in the Fourier domain:
\begin{eqnarray}
\label{HDM}
\sum'_{\mathbf{q}} \left(\begin{matrix}
c^\dagger_{\mathbf{q},1} && c^\dagger_{\mathbf{q},2}
\end{matrix}\right)\left(\begin{matrix}
\Delta(\mathbf{q}) && f(\mathbf{q})^\dagger +\delta(\mathbf{q})^\dagger \\ f(\mathbf{q})+\delta(\mathbf{q}) && -\Delta(\mathbf{q})
\end{matrix}\right) \left(\begin{matrix}
c_{\mathbf{q},1} \\ c_{\mathbf{q},2}
\end{matrix}\right)
\end{eqnarray}
In this equation, $\sum'_{\mathbf{q}}$ runs over half of the Brillouin zone to avoid double counting (as $c^\dagger_{\mathbf{ q}}=c_{\mathbf{-q}}$). Here $\delta(\mathbf{q})$ is a correction which changes position of conic points $\mathbf{Q}_{1,2}$ and modifies period of the spatial oscillations for the magnetic field-induced spin correlations \cite{long}. 
The gap is produced by the diagonal terms in Eq. (\ref{HDM}):
\begin{eqnarray}
\Delta(\mathbf{q})=2(\Gamma_{zx}\sin((\mathbf{q},\mathbf{n}_1))-\Gamma_{yz}\sin((\mathbf{q},\mathbf{n}_2))+\Gamma_{xy}\sin((\mathbf{q},\mathbf{n}_2-\mathbf{n}_1))).
\label{Delta}
\end{eqnarray}
The spectrum of this system is $\xi(\mathbf{q})=\sqrt{|f(\mathbf{q})|^2+\Delta(\mathbf{q})^2}$. In the vicinity of conic points, it turns out to be:
\begin{equation}
\varepsilon(\mathbf{Q_{1,2}+\delta q})=\sqrt{3K^2(\delta q)^2+\Delta^2(\mathbf{Q_{1,2}})},
\end{equation}
where  $\Delta(\mathbf{Q}_2)=-\Delta(\mathbf{Q}_1)=\sqrt{3}(\Gamma_{xy}+\Gamma_{yz}+\Gamma_{zx})=\sqrt{3}(\mathbf{h},\mathbf{D})(\tilde{A}+2\tilde{B}+\tilde{C})=\Delta$ with $\mathbf{h}=(h^x,h^y,h^z)$, $\mathbf{D}=(D^x,D^y,D^z)$. 
 
The gap in the spectrum renders decay of spin correlation functions at large distance $r\gg l=\frac{|\Delta|}{\sqrt{3}K}$ exponential. In limit $r\gg l\gg t$ and $D^\alpha\gg h^\beta$, spin-spin correlation function becomes: 
\begin{eqnarray}
\label{Sab}
S^{\alpha\beta}_{ab}(\mathbf{r},t)=-\zeta_a \zeta_b \frac{3D^\alpha D^\beta(\tilde{A}-\tilde{C}+2\tilde{B})^2e^{-\frac{2 r}{l}}}{4\pi^2 l^2r^2},
\end{eqnarray} 
where $\zeta_a=1$ if $a$ is even and $\zeta_a=-1$ if $a$ is odd.
In Eq.(\ref{Sab}) we kept the main term only, assuming  $|\mathbf{h}| \ll |\mathbf{D}|$.

\section{Conclusions}

We developed a general method to analyse the infrared behaviour of spin-spin correlation functions
in the Kitaev honeycomb model with various types of local perturbations, producing power-law  behaviour of
correlations.  We demonstrated that calculation of spin-spin correlation functions can be reduced to the calculation
of some  specific correlation functions of bilinear forms of Majorana fermions, described by quadratic Hamiltonian.
In the lowest order over the perturbation strength, the Majorana Hamiltonian coincides with the bare one derived 
originally by Kitaev. Higher-order terms in the expansion over the various perturbations can be accounted for by
the corrections to the effective Majorana Hamiltonian we derived. Properties of the perturbations with respect to time inversion have been used to explain qualitative differences between the results 
previously obtained in the theory of Kitaev model with different types of perturbations in Refs.~\onlinecite{TFK2011,PhysRevLett.117.037209}.

Our specific new results refer to the Kitaev model with the perturbation of the DM type that was
predicted to exist in some relevant magnetic  compounds, like $\alpha$-RuCl$_3$ and especially $\alpha$-Li$_2$IrO$_3$.
We demonstrated that DM perturbation leads to the power-law correlations already in the lowest possible (second)
order over its strength $D/K$, and calculated explicit form of the spin-spin correlation functions in real space-time,
as well as the corresponding spin structure factor in $\omega-\mathbf{q}$ representation.
This result indicates a primary importance of the DM perturbation among other possible type of perturbations,
since results of Ref.~\onlinecite{PhysRevLett.117.037209} show that Heisenberg  ($J$) and pseudodipolar 
($\Gamma$) perturbations produce long-range correlations in the  order $J^4\Gamma^4/K^8$ only.
We have also studied the result of  application of external magnetic field $\mathbf{h}$ together with DM
perturbation, and found that it produces a gap in the Majorana fermion spectrum; the magnitude of this
gap $\Delta \sim D h/K$ grows linearly with $h$; similar behaviour was recently observed experimentally in 
Refs. \onlinecite{PhysRevLett.119.037201,hentrich2017large}.

KT gratefully acknowledges the financial support of the Ministry
of Education and Science of the Russian Federation in the framework of Increase Competitiveness
Program of MISIS. This research was also partially supported by the RF Presidential Grant No. NSh-10129.2016.2.
We are grateful to A.Yu.Kitaev and A.A.Tsirlin for useful discussions.

\bibliographystyle{apsrev}
\bibliography{bibliography}
\end{document}